\documentclass[runningheads]{llncs}
\usepackage{multirow}
\usepackage[table,xcdraw]{xcolor}
\usepackage{enumerate}
\usepackage{url}
\usepackage{adjustbox}

\usepackage{cite}
\usepackage{amsmath,amssymb,amsfonts}
\usepackage{algorithmic}
\usepackage{algorithm}
\usepackage{xcolor}
\usepackage{graphicx}
\usepackage{textcomp}
\usepackage{ragged2e}
\usepackage{marvosym}
\begin{document}
\mainmatter 
\setcounter{secnumdepth}{4}
\title{Retrieval-augmented GPT-3.5-based Text-to-SQL Framework with Sample-aware Prompting and Dynamic Revision Chain}
\titlerunning{ICONIP23} 
%

\author{Chunxi Guo, Zhiliang Tian\textsuperscript{ (\Letter)}, Jintao Tang, Shasha Li, Zhihua Wen, \\Kaixuan Wang and Ting Wang\textsuperscript{ (\Letter)}
}
\authorrunning{Guo et al.} 
\institute{College of Computer, National University of Defense Technology, Changsha, China\\
\email{\{chunxi, tianzhiliang, tangjintao, shashali, zhwen, wangkaixuan18, tingwang\}@nudt.edu.cn}\\
}
\maketitle 

\begin{abstract}
Text-to-SQL aims at generating SQL queries for the given natural language questions and thus helping users to query databases. 
Prompt learning with large language models (LLMs) has emerged as a recent approach, which designs prompts to lead LLMs to understand the input question and generate the corresponding SQL. 
However, it faces challenges with strict SQL syntax requirements.
Existing work prompts the LLMs with a list of demonstration examples (i.e. question-SQL pairs) to generate SQL, but the fixed prompts can hardly handle the scenario where the semantic gap between the retrieved demonstration and the input question is large. 
In this paper, we propose a retrieval-augmented prompting method for a LLM-based Text-to-SQL framework, involving sample-aware prompting and a dynamic revision chain.
Our approach incorporates sample-aware demonstrations, which include the composition of SQL operators and fine-grained information related to the given question.
To retrieve questions sharing similar intents with input questions, we propose two strategies for assisting retrieval.
Firstly, we leverage LLMs to simplify the original questions, unifying the syntax and thereby clarifying the users' intentions.
To generate executable and accurate SQLs without human intervention, we design a dynamic revision chain which iteratively adapts fine-grained feedback from the previously generated SQL.
Experimental results on three Text-to-SQL benchmarks demonstrate the superiority of our method over strong baseline models.

\end{abstract}

\section{Introduction}
Text-to-SQL task aims to convert natural language question (NLQ) to structured query language (SQL), allowing non-expert users to obtain desired information from databases \cite{ratsql,Cai_Xu_Zhang_Yang_Li_Liang_2018_Encoder-Decoder}.
As databases are popular in various scenarios involving different domains (e.g., education and financial systems, etc.), it is desirable to train a model that generalizes well across multiple domains. To facilitate cross-domain generalization \cite{picard,SADGA}, researchers adapt encoder-decoder architecture \cite{li2023resdsql,Graphix-T5}, reducing the requirement for specific domain knowledge via end-to-end training. These approaches require diverse and extensive training data to train the model, which is prohibitively expensive~\cite{zhao2023survey}.

Recent progress focuses on large language models (LLMs) (e.g., GPT-3~\cite{GPT3}, Codex~\cite{Codex} and GPT-4~\cite{openai2023gpt4}) with prompt learning~\cite{liu2023pl}, which refers to using specific prompts or instructions to generate desired responses.
Rajkumar et al.~\cite{EvaluatingLLM} and Liu et al.~\cite{Evaluatingchatgpt} evaluate several prompt learning baselines for Text-to-SQL tasks.
Their findings show that though it is natural for LLMs to generate text sequences, generating SQL is still a challenge due to the SQL's strict syntax requirements.
To address these issues, inspired by few-shot learning~\cite{liu2023pl}, existing work employs prompting the LLMs with a list of demonstration examples (i.e. question-SQL pairs) to generate SQL queries.
However, they typically rely on manual labour to create static demonstration examples tailored to specific tasks.
DIN-SQL~\cite{pourreza2023dinsql} selects pre-defined samples from each category, 
SELF-DEBUGGING~\cite{chen2023teaching} explains the code to LLMs but without explanation demonstration.
These methods employ a static demonstration, meaning that the demonstration examples provided to LLMs are fixed and do not adapt or change across different examples. 
These static demonstration examples hardly adapt to the scenarios where the semantic gap between retrieved demonstrations and the input question is large, which is called retrieval bias \cite{Song}, commonly appearing in the retrieval-augmented generation.
Inspired by \cite{min2022rethinking}, we argue that providing dynamic demonstrations can be adaptive to specific samples and schema for SQL generation. 
Dynamic examples enable the SQL generation to accommodate various scenarios. By adjusting to specific instances, demonstrations can be customized to incorporate the necessary query structure, logical operations, and question semantics. This adaptability facilitates the SQL generation that are relevant and appropriate for different situations.

In this paper, we propose retrieval-augmented prompts for an LLM-based Text-to-SQL model, which contains sample-aware prompting and a dynamic revision chain.
Specifically, we propose to retrieve similar SQL queries to construct prompts with sample-aware demonstration examples. 
Notice that users often ask questions in different expressions, even if they have the same intention and SQL query. It makes the model hard to retrieve helpful examples.
To solve this issue, we propose to extract the question's real intention via two strategies:
Firstly, we simplify original questions through LLMs to clarify the user's intentions and unify the syntax for retrieval.
Secondly, we extract question skeletons for retrieving items with similar question intents. 
To produce executable and accurate SQL, we design a dynamic revision chain, generating SQL queries by iteratively adapting to fine-grained feedback according to the previous version of generated SQL. The feedback includes SQL execution results, SQL explanations, and related database contents.
This dynamic chain manages to generate executable and accurate SQL through automatic interaction between the language model and the database without human intervention.

Our contributions are as follows: 
(1) We develop a retrieval-augmented framework for Text-to-SQL tasks by prompting LLMs with sample-aware demonstrations.
(2) We propose a dynamic revision chain, which adapts to the previously generated SQL with fine-grained feedback.
(3) Experimental results on three Text-to-SQL benchmarks show that our method surpasses the strong baseline models.

\section{Related Work}

\subsection{Encoder-Decoder SQL Generation.} 
SQL generation tasks have achieved significant advancements through the utilization of encoder-decoder architectures \cite{Cai_Xu_Zhang_Yang_Li_Liang_2018_Encoder-Decoder}.

On the encoder side, Guo et al. \cite{guo2019complex} proposed IRNET, using attention-based Bi-LSTM for encoding and an intermediate representation-based decoder for SQL prediction. Later, \cite{Bogin_Berant_Gardner_2019,Chen_Chen_Zhao_Cao_Xu_Zhu_Yu_2021_ShadowGNN} introduced graph-based encoders to construct schema graphs and improve input representations.
Works such as RATSQL \cite{ratsql}, SDSQL \cite{li2023resdsql}, LGESQL \cite{LGESQL}, $S^{2}SQL$ \cite{hui-etal-2022-s2sql}, $R^{2}SQL$ \cite{Hui_Geng_Ren_Li_Li_Sun_Huang_Si_Zhu_Zhu_2021}, SCORE \cite{yu2021score}, and STAR \cite{STAR_Cai_2022} further improved structural reasoning by modelling relations between schemas and questions.
GRAPHIX-T5 \cite{Graphix-T5} overcomes the limitations of previous methods by incorporating graph representation learning in the encoder. Concurrently, RASAT \cite{rasat} also provided T5 with structural information by adding edge embedding into multi-head self-attention.

On the decoder side, we divide the methods into four categories: sequence-based methods (BRIDGE \cite{lin2020bridging}, PICARD \cite{picard}) directly translate NLQ into SQL query token by token, template-based methods (X-SQL \cite{X-SQL}, HydraNet \cite{hybrid}) employ predefined templates to regulate SQL generation and ensure structural coherence, stage-based methods (GAZP \cite{GAZP}, RYANSQL \cite{choi-etal-2021-ryansql}) first establish a coarse-grained SQL framework and then fills in the missing details in the frame which calls slot-filling methodologies, and hierarchical-based methods (IRNet \cite{irnet}, RAT-SQL \cite{ratsql}) generate SQL according to grammar rules in a top-down manner, resulting in a tree-like structure. 

\subsection{LLM-based SQL Generation.} LLM-based models recently emerge as a viable option for this task~\cite{EvaluatingLLM,Evaluatingchatgpt}.
For effectively utilizing, it is important to design appropriate in-context demonstration~\cite{brown2020language} and chain-of-thought (CoT)~\cite{wei2023chainofthought} strategies that can elicit its ability~\cite {zhao2023survey}.

In terms of searching for sample demonstration examples, DIN~\cite{pourreza2023dinsql} selects a fair number of demonstration examples from each category (e.g. simple classes, non-nested complex classes and nested complex classes), but they are fixed.
Moreover, Guo et al.~\cite{guo2023casebased} adaptively retrieve intention-similar SQL demonstration examples through de-semanticization of the questions. However, none of these methods can solve the ambiguous and varied questioning of realistic scenarios.

As for the CoT prompting strategy, DIN-SQL~\cite{pourreza2023dinsql} follows a least-to-most~\cite{zhou2023leasttomost} prompting method, decomposing Text-to-SQL task into subtasks and solves them one by one. Pourreza and Chen et al. explore self-correction~\cite{pourreza2023dinsql, chen2023teaching}, where the LLM explain the question and SQL, providing valuable feedback for improvement. Tian et al.~\cite{tian2023interactive} propose interactive generation with editable step-by-step explanations, combining human intervention with LLM generation to refine the final SQL output. Additionally, Sun et al. ~\cite{sun2023sqlpalm} explore execution-based self-consistent prompting methods.

Nonetheless, creating task-specific demonstration examples~\cite{pourreza2023dinsql, chen2023teaching,tian2023interactive,sun2023sqlpalm} demands manual labour. Instead, ours works through automatic interaction between the LLMs and the databases without human intervention.
Moreover, explaining to itself and simple feedback alone~\cite{chen2023teaching,pourreza2023dinsql} are weak for digging out errors for correction. While our approach takes into account all three aspects of fine-grained feedback, which interact with each other to create effective feedback.

\section{Methodology}
\begin{figure}[t] \centering
 \includegraphics[width=1.0\textwidth]{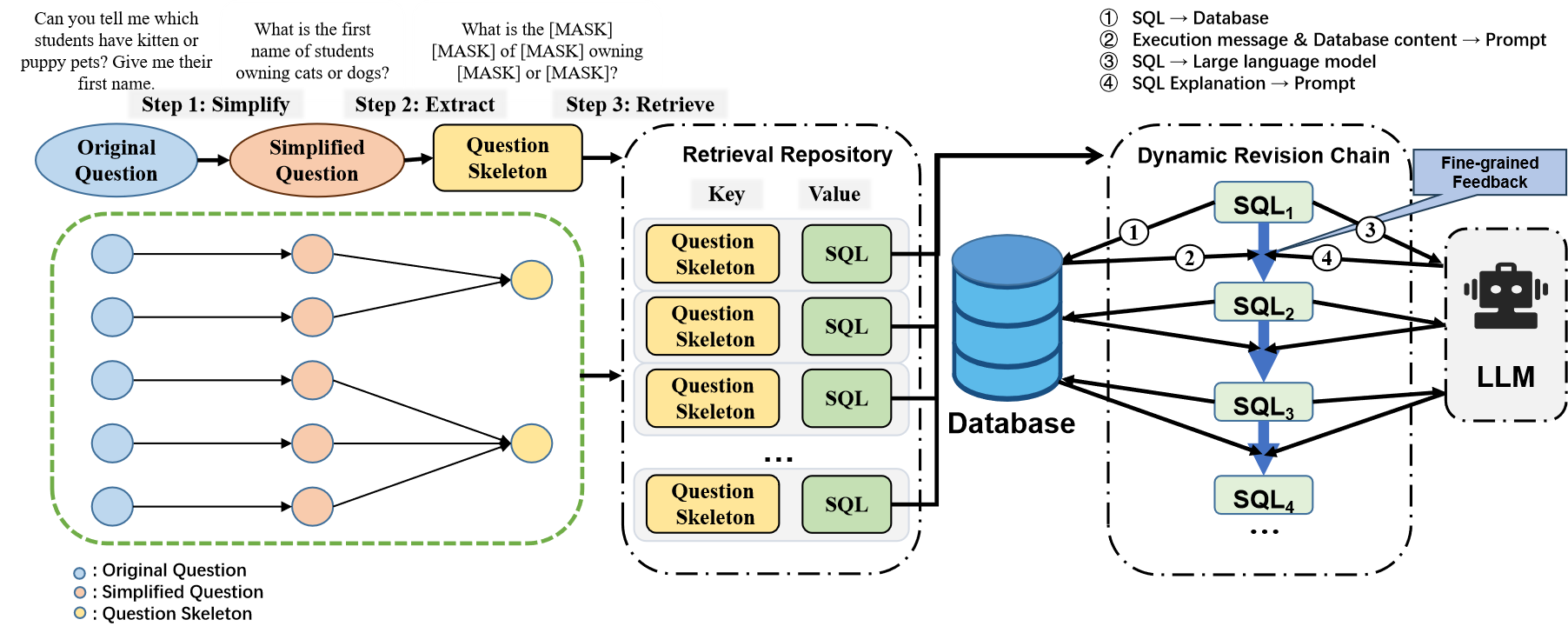}
 \caption{Framework overview: The left half shows retrieval repository construction in three steps. The top three sentences are three specific instances each. The Green dashed box presents the training set. The right half is a dynamic revision chain with SQL queries generated by LLM iterations as nodes (green boxes). The output of steps 2 and 4 are collectively referred to as fine-grained feedback.} \label{fig:model}
\end{figure}

Our framework consists of two modules as shown in Fig.~\ref{fig:model}: 
(1) \textbf{Retrieval Repository:} (see Sec.~\ref{Simplification and Retrieval}) We construct a retrieval repository with simplified questions added and then use question skeletons to retrieve sample-aware SQL demonstration examples.
(2) \textbf{Dynamic Revision Chain:} (see Sec.~\ref{Dynamic Revision Chain}) We further revise the generated SQL queries by adding fine-grained feedback.

\subsection{Retrieval Repository}
\label{Simplification and Retrieval}
We construct a retrieval repository consisting of multiple key-value retrieval items, where the keys represent the question skeletons and the values are $k$ sample-aware SQL queries. These processes enable us to generate demonstration examples that showcase desired behaviours of the LLM.
Our method involves:
(1) Simplifying original questions to unify various questioning styles (see Sec.~\ref{Question Simplification}).
(2) Extracting question skeletons to construct a retrieval repository (see Sec.~\ref{Skeleton Extraction}).
(3) Retrieving SQL queries according to skeleton similarities (see Sec.~\ref{Sample Retrieval}).

\subsubsection{Question Simplification.}
\label{Question Simplification}
We simplify natural language questions by prompting the LLM with instructions.
In this way, we can avoid the frustration of unusual questioning styles and enhance the syntax and wording variety in the repository.

Specifically, we construct a prompt template $prompt(.)$: \textit{``Replace the words as far as possible to simplify the question, making it syntactically clear, common and easy to understand: [QUESTION]"}, where ``\textit{[QUESTION]}'' represents the original natural language question.
We then obtain the simplified question by feeding $prompt(Q)$ into the LLM.
We maintain a consistent temperature setting in the language model to ensure that all simplified sentences exhibit the same probability distribution.

\subsubsection{Question Skeleton Extraction.}
\label{Skeleton Extraction}
We then extract question skeletons, including both original questions and simplified questions. We follow the method proposed by Guo et al.~\cite{guo2023casebased} to obtain question skeletons. This process removes specific schema-related tokens from the questions, focusing solely on the structure and intent. Finally, we take the (question skeleton, SQL) pairs from the training set and store them in the retrieval repository. Note that the number of samples in the retrieval repository is twice as large as the training set, due to the addition of the simplified samples.

Let $\mathcal{D}_{\text{train}}$ represents the training set, and $R$ denotes the retrieval repository. The original natural language question is denoted as $Q_{\text{o}}$, while $Q_{\text{r}}$ represents the simplified question. The question skeletons are denoted as $S_{\text{o}}$ and $S_{\text{r}}$ for the original and simplified questions, respectively. We formalize the composition of the retrieval repository as follows: $$R = \{(S_{\text{o}}, \text{SQL}), (S_{\text{r}}, \text{SQL}) \mid (Q_{\text{o}}, \text{SQL}) \in \mathcal{D}_{\text{train}}\}.$$


\subsubsection{Sample Retrieval.}
\label{Sample Retrieval}
The retrieval process searches for the most similar question skeletons and returns their corresponding SQL queries from the retrieval repository.
This search is based on the semantic similarity between the skeleton of the new question and the items' keys in $R$.

Specifically, given a new question $\widetilde{Q_{\text{o}}}$, we first obtain its simplified sentence $\widetilde{Q_{\text{r}}}$, and their corresponding question skeletons $\widetilde{S_{\text{o}}}$ and $\widetilde{S_{\text{r}}}$, following the same method used in previous two subsections (see \ref{Question Simplification} and \ref{Skeleton Extraction}). 
And then we calculate the cosine similarity scores $s_o$ between the semantic vector of question skeleton $\widetilde{S_{\text{o}}}$ and all questions skeletons $S$ in $R$. Similarly, we also compute the cosine similarity scores $s_r$ for simplified question skeleton $\widetilde{S_{\text{r}}}$ using the formula: $s = \cos \left( \mathbf{f}(S) \cdot \mathbf{f}(\widetilde{S})\right)$,
where $\mathbf{f}(.)$ represents an off-the-shelf semantic encoder\footnote{We utilize SBERT~\cite{sbert} in our experiment.}. Here, $\widetilde{S}$ will be instantiated as $\widetilde{S_{\text{o}}}$ and $\widetilde{S_{\text{r}}}$, and $s$ will be instantiated as $s_o$ and $s_r$, correspondingly.

From these scores, we select the top-$k$ retrieval samples with the highest rankings.
Let $k_1$ and $k_2$ denote the number of samples retrieved from the original question skeleton $\widetilde{S_{\text{o}}}$ and the simplified question skeleton $\widetilde{S_{\text{r}}}$ respectively, such that $k = k_1 + k_2$. We then concatenate the $k$ samples to form a demonstration example as input to the LLM.
Our retrieval repository offers the LLM with sample-aware SQL demonstration examples, which display a more practical answer space.

\subsection{Dynamic Revision Chain}
\label{Dynamic Revision Chain}
We employ the LLM to generate an initial SQL query, and then we iteratively revise the generated SQL queries based on fine-grained feedback, forming a dynamic revision chain.
The dynamic revision chain consists of: the SQL queries generated by the LLM iteration as nodes and the prompts provided to the LLM as edges.
It helps to generate executable and accurate SQL queries through interaction between language models and databases, with minimal human intervention.
The dynamic revision chain contains two stages: (1) assembling prompt based on the fine-grained feedback (see Sec.~\ref{Fine-grained Feedback}), and (2) generating SQL via iterative prompting (see Sec.~\ref{Iterative SQL Generation}).

\subsubsection{Fine-grained Feedback.}
\label{Fine-grained Feedback}
We collect three fine-grained pieces of information based on the SQL generated in the previous iteration.
The intuition is that various information hampers LLMs' focus, so they struggle to extract necessary data from extensive and complex databases. Thus, we should progressively narrow down the scope and prioritize the most likely information.
The fine-grained feedback in our approach consists of three aspects of information:

\begin{enumerate}[(1)]
\item \textbf{Execution Error Feedback}: 
We feed the SQL query generated by LLM into the database engine (i.e. SQLite) for execution.
We then obtain the error messages reported during the execution and add them to the prompt.
It checks whether the predicted SQL can be executed correctly, and report the specifics of the error (e.g. \textit{``no such table: [TABLE]"}, \textit{``no such function: YEAR}, \textit{``misuse of aggregate: COUNT()"}).
By incorporating the execution error messages into the prompt, LLM can learn from its errors. This helps to generate queries that follow the SQL syntax rules.

\item \textbf{Natural Language Explanation}: 
We prompt the LLM with instructions, converting the SQL predicted in the previous iteration back into its corresponding natural language expression. 
Specifically, we construct an instruction:\textit{``What does this SQL query mean? What are the differences between the predicted meaning and the question meanings above?"}
The LLM identifies semantic gaps and fills them by explaining the meaning of its own generated SQL and comparing it to the meaning of the original question.

\item \textbf{Related Database Contents}: 
We provide the LLM with content details about the database tables and columns involved in the SQL queries predicted in the previous iteration, including the possible values involved in the question. It aims to allow LLMs to simulate execution and thus generate more contextually relevant and accurate SQL queries.
\end{enumerate}

Overall, the fine-grained feedback approach aims to enable LLMs to learn from their mistakes, understand the meaning of the SQL queries generated and use contextual information in the database to generate more accurate and relevant SQL queries. By addressing challenges and focusing on important aspects, the methodology aims to help the LLm better extract the necessary data from complex databases and improve the performance of its query generation.

\subsubsection{Iterative SQL Generation.}
\label{Iterative SQL Generation}
Based on prompts with fine-grained feedback, the LLM iteratively generates SQL queries.
The intuition for iterative generation is that one iteration of fine-grained feedback might not check for all mistakes, whereas multiple iterations of feedback generation are more likely to get progressively closer to the gold answer.

Specifically, we concatenate three fine-grained feedback components with the previously generated SQL in each iteration, feeding them into the LLM. We then obtain a new SQL and collect new fine-grained feedback based on it, proceeding so to iterative generation. Let's denote the previous SQL query generated by the LLM as $SQL_{prev}$ and the current SQL query as $SQL_{curr}$. The fine-grained feedback components are represented as $F_{error}$ for execution error feedback, $F_{NL}$ for natural language explanation, and $F_{DB}$ for related database contents.
At each iteration $i$, the LLM generates a new SQL query $SQL_{curr}^{(i)}$ by incorporating the fine-grained feedback components:
$$
SQL_{curr}^{(i)} = \text{{LLM}}(SQL_{prev}, F_{error}^{(i)}, F_{NL}^{(i)}, F_{DB}^{(i)}).
$$

After executing $SQL_{curr}^{(i)}$ using the database engine, we obtain the result $R_{prev}^{(i)}$ from the previous iteration and $R_{curr}^{(i)}$ from the current iteration. To avoid infinite loops, we set a maximum number of iterations $N_{max}$.
The termination condition is defined as: 
$R_{prev}^{(i)} = R_{curr}^{(i)}$ or $i = N_{max}$.
This control mechanism ensures that the generated SQL queries converge to an optimal and executable solution within a reasonable timeframe.




In this iterative feedback loop, we enable a dynamic interaction between the LLM and the database engine, maximizing the generation of executable SQL without extensive human intervention.

\section{Experiments}

\subsection{Experimental Setup}

\subsubsection{Setting.}

We evaluate our method on text-davinci-003, which offers a balance between capability and availability. We apply FAISS \cite{FAISS} for storing the question skeletons and efficient retrieval followed by Guo et al.~\cite{guo2023casebased}. 
For the initial simplification of questions, we set temperature $\tau$=1.0. When generating SQL samples, we set temperature $\tau$=0.5.
For the number of retrieval samples, we assign $k_1$=4 and $k_2$=4.

\subsubsection{Datasets.}
We conduct experiments on the cross-domain large-scale Text-to-SQL benchmark as follows: (1) \textbf{Spider} \cite{spider} is a large-scale benchmark of cross-domain Text-to-SQL across 138 different domain databases. (2) \textbf{Spider-Syn} \cite{syn} is a challenging variant based on Spider that eliminates explicit alignment between questions and database schema by synonym substitutions. (3) \textbf{Spider-DK} \cite{dk} is also a variant dataset based on Spider with artificially added domain knowledge.

\subsubsection{Evaluation.}
We consider three key metrics: execution accuracy (EX) and test-suite accuracy (TS) \cite{ts}. EX measures the accuracy of the execution results by comparing them with the standard SQL queries, while TS measures whether the SQL passes all EX evaluations for multiple tests, generated by database augmentation.

Noting that EX is the most direct indication of the model performance in Text-to-SQL, although it contains false positives. Exact match evaluation is not performed, as multiple correct SQLs exist for one query. We use the official TS evaluation procedure, while for EX, we slightly modify the evaluation procedure due to the need to decouple the fine-tuning-based models for independent evaluation.


\subsubsection{Baselines.} We compare to two groups of methods:

\textbf{Fine-tuning T5-3B baselines:} \textbf{PICARD} \cite{picard} is a technique that constrains auto-regressive decoders in language models through incremental parsing; \textbf{RASAT} \cite{rasat}, which incorporates relation-aware self-attention into transformer models while also utilizing constrained auto-regressive decoders; and \textbf{RESDSQL} \cite{li2023resdsql}, which introduces a ranking-enhanced encoding and skeleton-aware decoding framework to effectively separate schema linking and skeleton parsing.

\textbf{Prompting LLMs baselines:} As for the large language models, we use two variants of the Codex family~\cite{Codex}~\cite{EvaluatingLLM} (\textbf{Davinci} and \textbf{Cushman}), \textbf{PaLM-2}~\cite{anil2023palm}~\cite{sun2023sqlpalm}, the \textbf{GPT-4} model~\cite{openai2023gpt4}~\cite{pourreza2023dinsql} and the \textbf{ChatGPT} model~\cite{Evaluatingchatgpt}.
In addition to a simple baseline assessment model, we choose several recent LLM-based work.
\textbf{DIN}~\cite{pourreza2023dinsql} decompose the Text-to-SQL tasks into sub-tasks: schema linking, query classification and decomposition, SQL generation, and self-correction; then performing few-shot prompting with GPT-4~\cite{openai2023gpt4}. 
\textbf{SELF-DEBUGGING}~\cite{chen2023teaching} adds error messages to the prompt and conducts multiple rounds of few-shot prompting for self-correction.
\textbf{Few-shot SQL-PaLM}~\cite{sun2023sqlpalm} adopts an execution-based self-consistency prompting approach. 


\subsection{Main Results}

\subsubsection{Performance on Spider Dataset.}
Table \ref{tab:main} displays the experimental outcomes of our proposed methods on Spider in comparison to the baseline methods. Across all three datasets, our methods achieve the highest levels of execution accuracy (EX) and test suite accuracy (TS).

Ours exhibits strong performance on test-suite accuracy, which exceeds the next-best method results in fine-tuning and prompting by 9.7\% and 5.9\% respectively.
In terms of execution accuracy, our method outperforms the next best method in both fine-tuning and prompting by 0.9\%.
For valid accuracy, our approach falls short of RESDSQL-3B + NatSQL and DIN-SQL (Few-shot) but also reaches 98.6\%. This is because prompting models may encounter challenges in generating SQL queries that adhere to strict syntactical and semantic rules, unless the individual steps and rules are as carefully designed as DIN~\cite{pourreza2023dinsql}.

\begin{table*}[]
\renewcommand\arraystretch{1}
\tabcolsep=0.2cm
\centering
\caption{Performance comparison on Spider with various methods. ``-" indicates that the results are not available. Schema indicates that the prompt contains the SQL for creating the database tables (i.e. tables, columns, value and its type)\protect\footnotemark[1]. }

\begin{tabular}{cccc}
\hline
\rowcolor[HTML]{C0C0C0} 
\textbf{Model} & \textbf{Method} & \textbf{EX} & \textbf{TS} \\ \hline
 & T5-3B + PICARD~\cite{picard} & 79.3 & 69.4 \\
 & RASAT + PICARD~\cite{rasat} & 80.5 & 70.3 \\
\multirow{-3}{*}{T5-3B} & RESDSQL-3B + NatSQL~\cite{li2023resdsql} & 84.1 & 73.5 \\ \hline
 & Few-shot~\cite{EvaluatingLLM} & 61.5 & 50.4 \\
\multirow{-2}{*}{Codex-cushman} & Few-shot + Schema~\cite{EvaluatingLLM} & 63.7 & 53.0 \\ \hline
 & Few-shot~\cite{EvaluatingLLM} & 60.8 & 51.2 \\
 & Few-shot + Schema~\cite{EvaluatingLLM} & 67.0 & 55.1 \\
 & Few-shot~\cite{pourreza2023dinsql} & 71.0 & 61.5 \\
 & DIN-SQL~\cite{pourreza2023dinsql} & 75.6 & 69.9 \\ 
\multirow{-5}{*}{Codex-davinci} & SELF-DEBUGGING~\cite{chen2023teaching} & 84.1 & - \\ \hline

& Few-shot SQL-PaLM~\cite{sun2023sqlpalm} & 82.7 & 77.3 \\ 
\multirow{-2}{*}{PaLM2} & Fine-tuned SQL-PaLM~\cite{sun2023sqlpalm} & 82.8 & 78.2 \\ \hline

& Zero-shot~\cite{pourreza2023dinsql} & 72.9 & 64.9 \\
& Few-shot~\cite{pourreza2023dinsql} & 76.8 & 67.4 \\
\multirow{-3}{*}{GPT-4} & DIN-SQL~\cite{pourreza2023dinsql} & 82.8 & 74.2 \\ \hline

ChatGPT & Zero-shot~\cite{Evaluatingchatgpt} & 70.1 & 60.1 \\ \hline
 & Zero-shot & 73.1 & 71.6 \\
\multirow{-2}{*}{Text-davinci} & Ours & \textbf{85.0 (0.9↑)} & \textbf{83.2 (5.9↑)} \\ \hline
\end{tabular}
\label{tab:main}
\end{table*}

\footnotetext[1]{In the Codex-davinci model, both methods utilize default with few-shot, but their demonstration examples are different.}


\textbf{Comparison with Zero-shot Prompting Models:}
On all three metrics, ours surpasses Codex, ChatGPT and even GPT-4 models with zero-shot prompting, even though they use the official format\footnote{https://platform.openai.com/examples/default-sqltranslate}. This indicates that although LLMs are trained using a specific format, their acquired capabilities are internalized and extended to use in more free formats.



\textbf{Comparison with Few-shot Prompting Models:}
Ours also outperforms all models in a few-shot setting. Notice that the two methods with similar few-shot prompting in the Codex-davinci model, the latter performs 10\% better than the former in both EX and TS. It indicates that the selection of demonstration examples (easy, non-nested complex, and nested complex classes)~\cite{pourreza2023dinsql} plays a significant role. While, ours uses an adaptive sample-aware sampling method brings 8.2\% more effective than this static demonstration suggests that incorporating more effective prompts is crucial for enabling the language model to understand new specific tasks.

\textbf{Comparison with Other Models:}
The closest EX performance compared to ours is SELF-DEBUGGIN, which also takes an iterative prompting strategy, but we still outperform it by 0.9\%.
Additionally, ours outperform SQL-PaLM, which uses a simpler prompting strategy but produces better results. It implies that PaLM2 is a potential LLM. We attempted to apply a consistency approach similar to that but failed. This may indicate that different LLMs are better suited to different approaches. Nevertheless, our adaptive sampling and iterative feedback approach has proven to be effective.

\begin{table*}[h]
\renewcommand\arraystretch{1}
\tabcolsep=0.3cm
\centering
\footnotesize
\caption{Evaluation of our method on Spider-SYN and Spider-DK datasets\protect\footnotemark[3]. ``-" indicates that the results are not available\protect\footnotemark[4].}
\begin{tabular}{cccc}
\hline
\multicolumn{4}{c}{\cellcolor[HTML]{D9D9D9}\textbf{SPIDER-SYN}} \\ \hline
\multicolumn{2}{c}{\textbf{Method}} & \textbf{EX} & \textbf{TS} \\ \hline
 & T5-3B + PICARD & 69.8 & 61.8 \\
 & RASAT + PICARD & 70.7 & 62.4 \\
\multirow{-3}{*}{Fine-tuning} & RESDSQL-3B + NatSQL & 76.9 & 66.8 \\ \hline
 & ChatGPT & 58.6 & 48.5 \\
 & Text-davinci & 60.7 & 60.3 \\
 & Few-shot SQL-Palm & 74.6 & 67.4 \\
 & Fine-tuned SQL-Palm & 70.9 & 66.4 \\
\multirow{-5}{*}{Prompting} & Ours & \textbf{81.4(4.5↑)} & \textbf{80.0(12.6↑)} \\ \hline
\multicolumn{4}{c}{\cellcolor[HTML]{D9D9D9}\textbf{SPIDER-DK}} \\ \hline
\multicolumn{2}{c}{\textbf{Method}} & \textbf{EX} & \textbf{TS} \\ \hline
 & T5-3B + PICARD & 62.5 & - \\
 & RASAT + PICARD & 63.9 & - \\
\multirow{-3}{*}{Fine-tuning} & RESDSQL-3B + NatSQL & 66.0 & - \\ \hline
 & ChatGPT & 62.6 & - \\
 & Text-davinci & 66.2 & - \\
 & Few-shot SQL-Palm & 66.5 & - \\
 & Fine-tuned SQL-Palm & 67.5 & - \\
\multirow{-5}{*}{Prompting} & Ours & \textbf{81.1 (13.6↑)} & - \\ \hline
\end{tabular}
\label{tab:syn-dk}
\end{table*}

\footnotetext[3]{As the other baseline models have not experimented on the Spider-SYN and Spider-DK datasets, there are relatively few models for comparison in the table.}
\footnotetext[4]{The Codex model could not be reproduced due to an invalid Codex'api. The TS metric was not applicable to the Spider-DK dataset.}

\subsubsection{Performance on Spider-SYN and DK Datasets.}
Table \ref{tab:syn-dk} showcases that our method demonstrates impressive robust performance compared with the baseline methods on Spider variants. On Spider-SYN, ours has 4.5\% and 12.6\% improved over the previous SOTA on EX and TS, respectively. Remarkably, ours performs surprisingly well with a 13.6\% improvement over the previous SOTA on Spider-DK.

\begin{table*}[]
\renewcommand\arraystretch{1}
\tabcolsep=0.18cm
\centering
\footnotesize
\caption{Test-suite accuracy at various complexity levels on Spider. The first two rows are standard few-shot prompting. }
\begin{tabular}{cccccc}
\hline
\textbf{Prompting} & \textbf{Easy} & \textbf{Medium} & \textbf{Hard} & \textbf{Extra} & \textbf{All} \\ \hline
Few-shot (CodeX-davinci)~\cite{pourreza2023dinsql} & 84.7 & 67.3 & 47.1 & 26.5 & 61.5 \\
Few-shot (GPT-4)~\cite{pourreza2023dinsql} & 86.7 & 73.1 & 59.2 & 31.9 & 67.4 \\
DIN-SQL (CodeX-davinci)~\cite{pourreza2023dinsql} & 89.1 & 75.6 & 58.0 & 38.6 & 69.9 \\
DIN-SQL (GPT-4)~\cite{pourreza2023dinsql} & 91.1 & 79.8 & 64.9 & 43.4 & 74.2 \\
Few-shot SQL-PaLM (PaLM2)~\cite{sun2023sqlpalm} & \textbf{93.5} & 84.8 & 62.6 & 48.2 & 77.3 \\
Fine-tuned SQL-PaLM (PaLM2)~\cite{sun2023sqlpalm} & \textbf{93.5} & 85.2 & 68.4 & 47.0 & 78.2 \\
Ours (Text-davinci) & 91.9 & \textbf{88.6} & \textbf{75.3} & \textbf{63.9} & \textbf{83.2} \\ \hline
\end{tabular}
\label{tab:level}
\end{table*}

\subsection{Various Difficulty Levels Analysis}
As shown in Table~\ref{tab:level}, we evaluate our effectiveness at various difficulty levels, which are determined by the number of SQL keywords used, the presence of nested sub-queries, and the utilization of column selections or aggregations.
The results show that ours outperforms the other models at all levels except for the easy level, where it is worse than SQL-PaLM.
The improvement in performance with increasing difficulty levels indicates that our model's strengths become more pronounced as the queries become more challenging.
This suggests that our model excels in handling complex SQL queries.

\subsection{Ablation Study}

\begin{figure}
\centering
\includegraphics[width=1.0\textwidth]{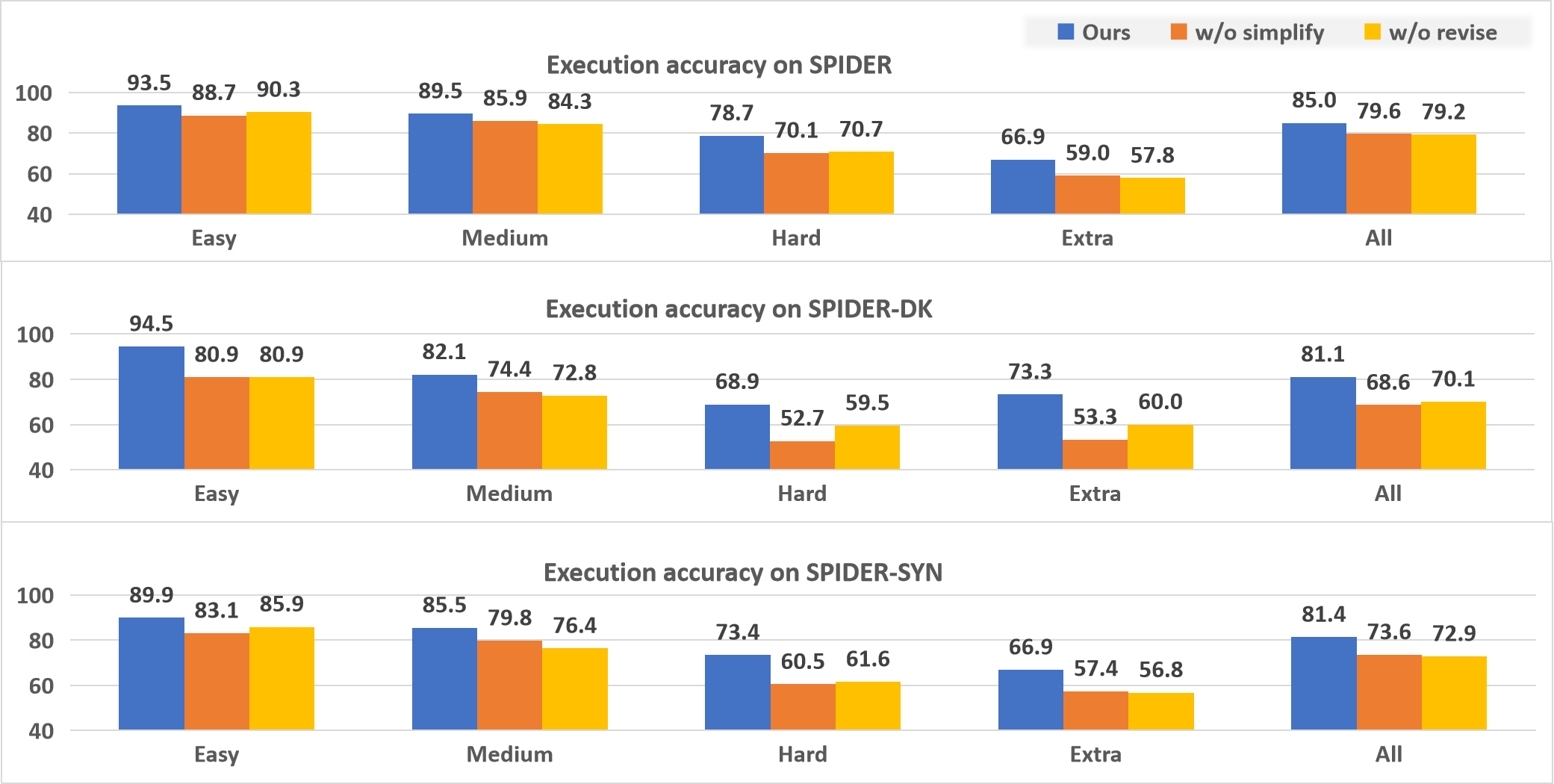}
 \caption{Ablation study of our model components at various complexity levels across three datasets. w/o simplify refers to using a direct retrieval of the question skeletons rather than a strategy of simplifying questions. w/o revise refers to removing the dynamic revision chain module.} \label{fig: ablation}
\end{figure}

Figure \ref{fig: ablation} demonstrates with and without each of the two modules at four complexity levels.
It shows that the exclusion of any of the modules leads to a decrease in performance at all levels of difficulty, in terms of hard and extra levels.
Decreases in model performance are similar for the w/o revise and w/o simplify settings.
Note that both modules of our method are most effective in improving the Spider-DK's easy level by 13.6\% each, which requires additional domain knowledge. This suggests that the simplification strategy and dynamic revision chain strategy contribute to a variety of generalisation issues.

We found that removing the simplification module resulted in a significant drop in model performance, particularly in the DK dataset where the overall drop was 12.5\%. The impact at different difficulty levels is in descending order of extra, hard, easy, and medium. This is possibly due to the fact that the model can incorporate more external knowledge as a supplementary description when simplifying, especially in the case of more SQL components. Note that w/o simplify is rather more effective for solving easy-level problems than medium-level ones, probably because the execution accuracy of easy-level problems is already high and short sentences are more likely to cause ambiguity.

Without the revision module, model performance suffers more as the difficulty level increases. On Spider-DK the model performance decreases by 11.0\%, especially on easy-level and extra-level by 13.6\% and 13.3\% respectively. As higher difficulty levels require more knowledge, this suggests that the fine-grained feedback in the revision module effectively complements the domain knowledge required for SQL generation. 

\begin{figure}
\centering
\includegraphics[width=0.8\textwidth]{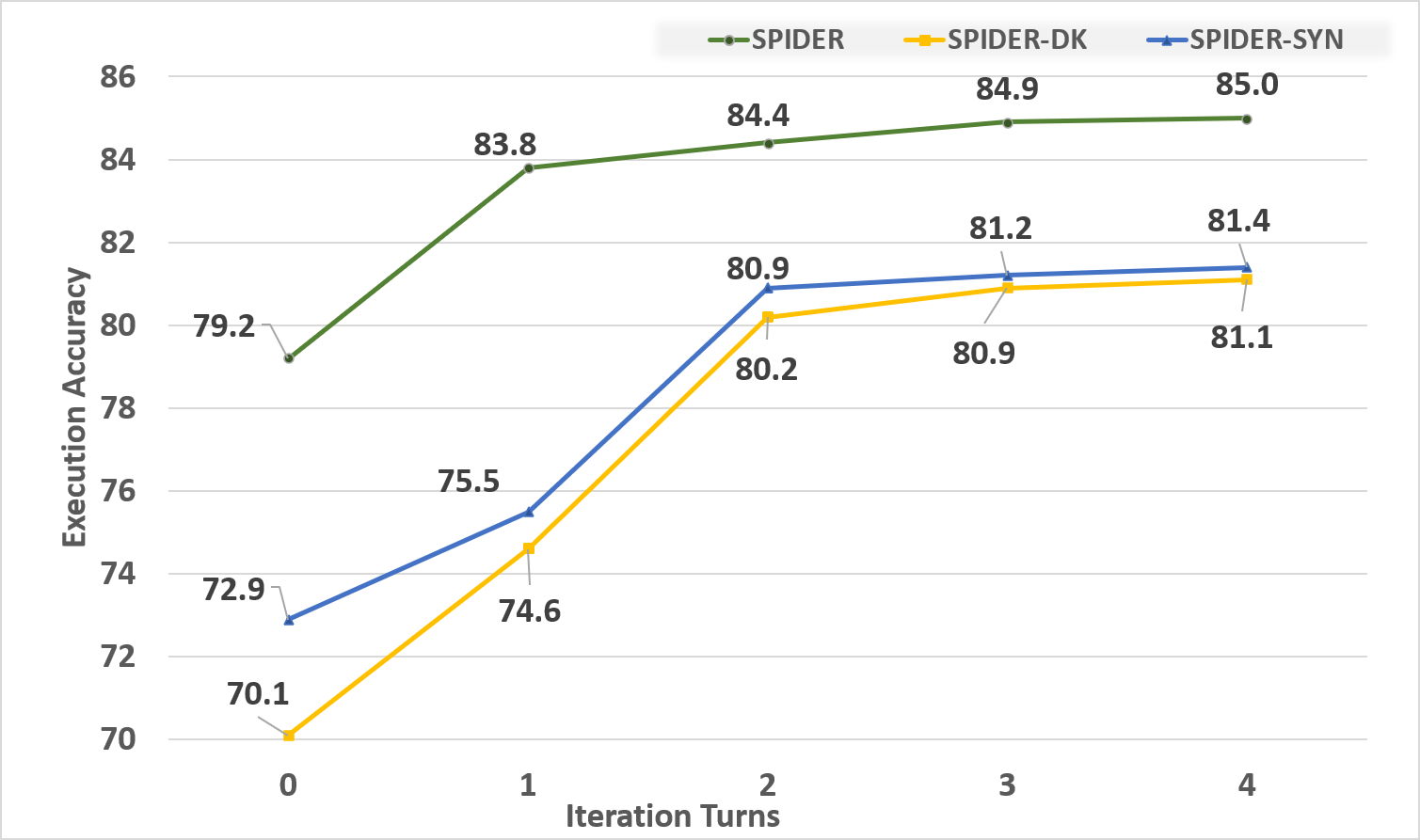}
 \caption{Analysis of dynamic SQL revision chain with different numbers of iteration rounds across three datasets: Spider, Spider-SYN, and Spider-DK. } \label{fig: iteration}
\end{figure}

\subsection{Iterative Round Analysis.}
From Fig. \ref{fig: iteration}, we observe that the major improvement comes from the first two iteration turns. Noting that in addition to the 4.6\% improvement in the first iteration of Spider, the other two datasets investigating generalizability, Spider-DK and Spider-SYN, showed a slightly better improvement in accuracy in the second iteration than in the first.
This indicates that iterative feedback of fine-grained information from a dynamic revision chain helps to deal with more complex generalisation problems, comparable to multiple reasoning needs to progressively derive the target answer.

\begin{figure}[H]
\centering
\includegraphics[width=1.0\textwidth]{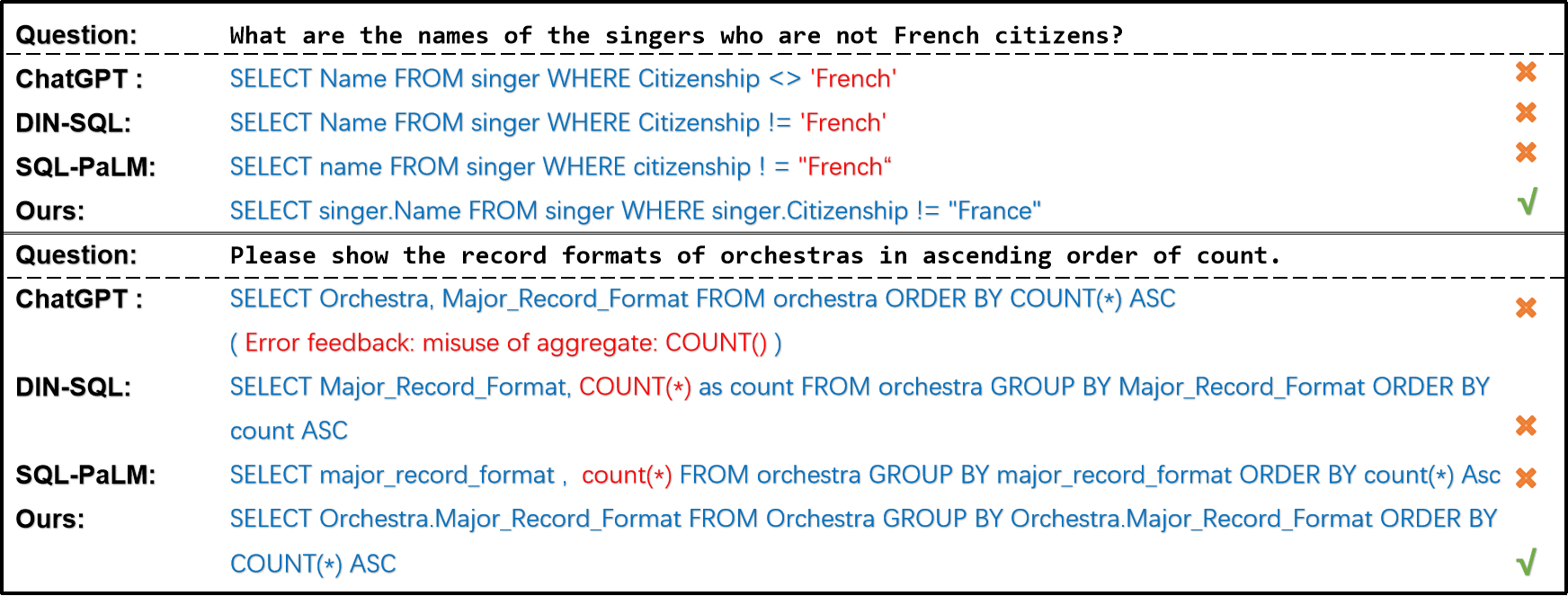}
 \caption{Two illustrative cases from Spider~\cite{spider}. Blue-coloured text is the correct generation, while the red-coloured text indicates the wrong generation. On the right hand side, $\checkmark$ means correct SQL while $\times$ means wrong.} \label{fig: case}
\end{figure}

\subsection{Case Study}
To demonstrate our model, we show a comparison of predicted SQL queries in Figure~\ref{fig: case} using ChatGPT~\cite{EvaluatingLLM}, DIN-SQL~\cite{pourreza2023dinsql}, SQL-PaLM~\cite{sun2023sqlpalm} and Ours.

In the first example, since the question obviously mentions \textit{``French"}, the general models will be confused about the exact value of the column \textit{``citizenship"} even if they pick it out. Noting that a SQL query must match the exact word mentioned to find the correct answer. Our approach provides the exact value of the database content involved in the first fine-grained iteration, which leads to a golden answer.

The second example requires only the selection of one item, whereas DIN-SQL and SQL-PaLM both select two. ChatGPT incorrectly uses the aggregate function \textit{COUNT()}, which in this case is required in conjunction with \textit{GROUP BY}. Our approach self-corrects the error in the second fine-grained iteration by interpreting the SQL interpretation in natural language.

\section{Conclusion}

We propose retrieval-augmented prompts for an LLM-based Text-to-SQL model.
By utilizing sample-aware prompting and a dynamic revision chain, we address the challenge of retrieving helpful examples and adapting the generated SQL based on fine-grained feedback.
Experimental results on three Text-to-SQL benchmarks demonstrate the effectiveness of our method.

%
\bibliographystyle{splncs03_unsrt} 
\bibliography{egbib.bib}
\end{document}